\documentclass[twoside,slac_one]{revtex4}
\usepackage{graphicx}
\usepackage{fancyhdr}
\usepackage{amsmath} % American Mathematics Society standards
\usepackage{bm}% bold math
\usepackage{amsxtra}
\usepackage{amssymb}
\usepackage{amsthm}
\usepackage{latexsym}
\usepackage{lscape}

\begin{document}

\title{
Studies of {\boldmath $b$}-hadron decays to charming final states at LHCb
}


\author{S. Ricciardi (on behalf of the LHCb Collaboration)}
\affiliation{STFC Rutherford Appleton Laboratory, Chilton, Didcot, Oxfordshire, OX11 0QX, UK}

\begin{abstract}
We present studies from the LHCb experiment of decays of the type $H_b \to H_c X$,
where $H_b$ represents a beauty hadron ($B^\pm$, $B^0$ or $\Lambda_b^0$) 
and  $H_c$  a charmed hadron ($D^0$, $D^{(*)+}$, $D_s^+$ or $\Lambda_c^+$).
Such decays are important for the determination of the CKM angle $\gamma$,
a key goal of the LHCb physics programme.
We exploit the data accumulated in 2010, and in the early months of the 2011
run.  We report on the observation of new decay modes, and first 
measurements on the road to a precise determination of $\gamma$.
\end{abstract}

\maketitle

\thispagestyle{fancy}


%%%%%%%%%%%%%%%%%%%%%%%%%%%%%%%%%%
\section{\label{introduction}Introduction}

Decays of $b$-hadrons (($H_b$) to open charm are of great interest both in the context of $CP$ violation studies and of QCD studies of heavy-quark dynamics. In particular, the angle $\gamma$ of the CKM Unitarity Triangle can be determined from $H_b\to (D^0,\bar{D^0})X_s$, where $D^0$ and $\bar{D^0}$ decay to a common final state, thanks to the interference of $b\to u$ and $b \to c$ tree-level transitions. 
In addition, the abundant and predictable relative rate of suitable decays to open-charm can be used to measure the production fractions of different $b$ meson and baryon species. Since all $b$-hadron species are produced in $pp$ collisions at the LHC, these measurements are needed 
to normalise measurements of $B^0_s$ and $\Lambda^0_b$ branching fractions to those of known $B^+$ or $B^0$ decays, and to determine absolute branching fractions.

In these proceedings, we report on the most recent results from LHCb using data accumulated in 2010, and in the early months of the 2011 run. In Section~\ref{charged}, we describe preliminary measurements on the road of a precise determination of $\gamma$  with charged $B$ decays. 
In Section~\ref{strange}, we present two different measurements of the $B_s$ production fraction, with semileptonic and fully hadronic decays, which combined give the most accurate determination of $f_s/f_d$. Finally, in Section~\ref{baryons}, we present the first observation of the $\Lambda^0_b\to D^0pK^-$ decay and a hint of the neutral beauty strange baryon $\Xi^0_b$, also reconstructed in the $D^0pK^-$ final state.

%%%%%%%%%%%%%%%%%%%%%%%%%%%%%%%%%%
\section{\label{experiment}The LHCb experiment}

The LHCb experiment has been designed to study decays of $b$-hadrons from $pp$ collisions at the LHC.
The detector has been described elsewhere~\cite{bib:LHCb}. Here, we just mention the salient experimental features which are critical for the measurement of $\gamma$ and are common to many hadronic decays to open-charm. Above all, since sensitivity to $\gamma$ arises from the interference of the $b\to c$ with the suppressed $b\to u$ amplitude, a large data sample is mandatory. It is ensured by: the high integrated luminosity that the LHC delivers, the large $b\bar{b}$ cross-section within the LHCb detector acceptance, and a dedicated and flexible trigger, which can select efficiently $b$-hadron decays. The vertex detector and the tracking system also play a crucial role: a momentum resolution smaller than 1\% and clear separation of secondary vertices from the primary vertex enable the separation of $b$-hadron decays from different sources of background components, both prompt,  from the primary vertex, and non-prompt, due to long-lived hadron decays other than signal. In addition, the excellent pion-kaon separation over a wide momentum range, provided by two RICH detectors, is vital to distinguish the different $H_b$ and $D$ decays of interest. For example, it is necessary to separate $B^+\to DK^+$, from the about ten times more abundant $B^+\to D\pi^+$ in the measurements of $\gamma$ using charged $B^+\to DK^+$ decays.\footnote{In the following, $D$ indicates a superposition of $D^0$ and $\bar{D^0}$.}

%%%%%%%%%%%%%%%%%%%%%%%%%%%%%%%%%%
\section{\label{charged}{ {\boldmath $CP$} violation studies with charged {\boldmath $B$} decays}}

Despite the impressive achievements by experiments at $B$-factories and the Tevatron, the CKM angle $\gamma$
is still the least well-determined angle of the Unitarity Triangle. The current average of direct measurements of $\gamma$ has an uncertainty of about $10^\circ$~\cite{bib:CKMUTfitters}. This precision can be significantly improved at LHCb in the near future.

Already now, the 2010 and early 2011 LHCb data-sets 
are sufficient to set constraints on the $CP$ asymmetries and measure the ratio of branching fractions of the most sensitive $B^+\to DK^+$ decay over the favoured $B^+\to D\pi^+$ mode. These measurements demonstrate the capability of LHCb in three well-established methods for extracting $\gamma$: the GLW method~\cite{bib:GLW}, which uses $D$ decays to $CP$-eigenstates (e.g., $K^+K^-$), the ADS method~\cite{bib:ADS}, where the $D^0$ or $\bar{D^0}$ is reconstructed in a final state accessible to both Cabibbo-favoured (CF) and doubly-Cabibbo-suppressed (DCS) transitions (e.g., $K^\pm\pi^\mp$), and the GGSZ method~\cite{bib:GGSZ}, which exploits the interference over the Dalitz plot of $D$ decays to three-body final states (e.g., $K^0_S\pi^+\pi^-$).

%In addition to these channels, which are the only ones effectively exploited by previous experiments, other decays sensitive to $\gamma$ are under study at LHCb. We will present preliminary  measurements for some of them in the following sections.

%%%%%%%%%%%%%%%%%%%%%%%%%%%%%%%%%%
\subsection{\label{glw}Measurements of GLW observables}

As first step towards a measurement of $\gamma$ with the GLW method, the ratio of the $B^+\to DK^+$ branching fraction to that of $B^+\to D\pi^+$ is measured using the 2010 LHCb dataset, corresponding to 36.5 pb$^{-1}$~\cite{bib:LHCbGLW}. The measurement is performed, simultaneously, for the Cabibbo-favoured (CF) $D\to K^+\pi^-$ and the $D\to K^+\pi^-\pi^+\pi^-$ decay, and, separately, for the $CP$-even $D \to K^+K^-$ (CP+) decay (a difference between the two can be expected due to the relative larger interference between the $b\to u$ and the $b \to c$ transitions in the $CP$-even case). 

\begin{figure*}[ht]
\centering
\includegraphics[width=135mm]{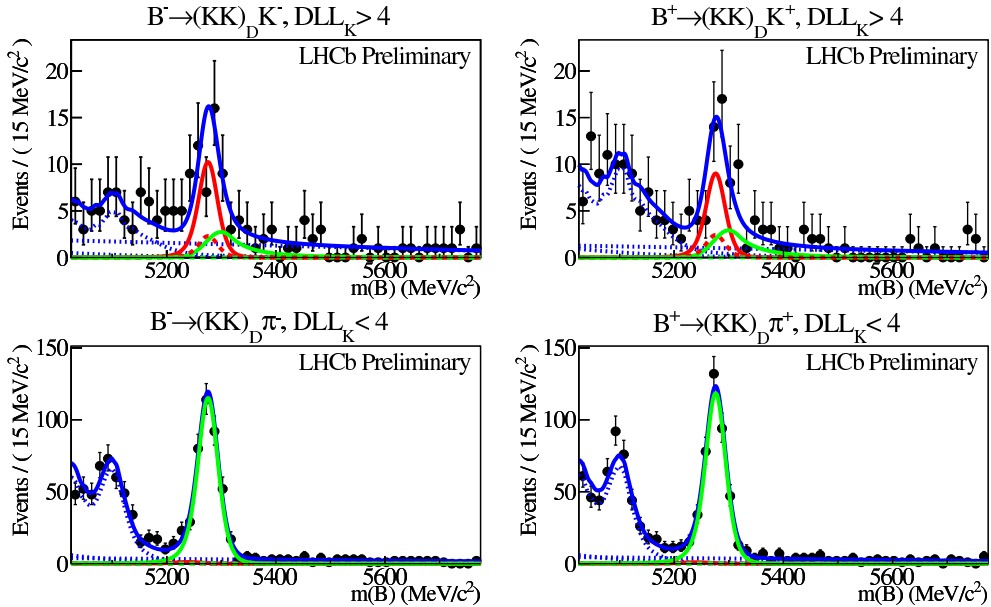}
\caption{Reconstructed $B$ mass distributions for $B^\pm \to (KK)_D K^\pm$ (top) and $B^\pm \to (KK)_D \pi^\pm$ (bottom) candidates. Sensitivity to charge asymmetry is obtained by splitting $B^-$ (left) and $B^+$ (right). The solid red (green) curve is the fitted $B^\pm\to DK^\pm$ ($B^\pm\to D\pi^\pm$) signal. The dashed lines indicate the different background components from: charmless (red and green, if present), combinatoric, partially reconstructed, and semileptonic decays (blue)\cite{bib:LHCbGLW}.    
} \label{fig:GLW}
\end{figure*}

The $B^+\to DK^+$ and  $B^+\to D\pi^+$ signal yields are extracted with an unbinned extended maximum-likelihood fit to the $B$-mass distributions. The fit is performed simultaneously to four different mass distributions, which are obtained by separating the sample according to the charge of the bachelor hadron, and the value of a particle identification discriminant for the bachelor. The used PID discriminant is the difference of the log-likelihood between the kaon and pion hypotheses, DLL$_{K\pi}$.
The results of the fit are shown in Fig.~\ref{fig:GLW} for the $D\to KK$ mode.

The ratio of branching fractions is computed from the fitted yields and the ratio of efficiencies. The PID efficiency determination uses a data calibration sample of pions and kaons from $D^{*+}\to D^0(K^-\pi^+)\pi^+$ decay and a re-weighting technique to take into account small difference in the kinematics between the calibration sample and the signal samples. Other efficiencies (geometric acceptance, trigger and reconstruction) are very similar for the two decay modes, and their ratio is derived from Monte Carlo simulations. 

The results are: 
${\cal{R}}_{CF}^{K/\pi} = (6.30 \pm 0.38 \pm 0.40)\%,$ and
${\cal{R}}_{CP+}^{K/\pi} = (9.31 \pm 1.89 \pm 0.53)\%.$  
From the ratio of the CP+ over CF measurements, the following $\gamma$-sensitive observable is computed:
$${\cal{R}}_{CP+} = 1.48 \pm 0.31 \pm 0.12.$$
%where the systematic uncertainty includes an additional 2.2\% theoretical systematic~\cite{bib:BABAR-theoryerror}.
In addition, three $CP$ asymmetries are measured between the $B^-$ and the $B^+$ decay rates:
\begin{eqnarray*}
{\cal{A}}_{CF~}^{DK} & = &(-0.08 \pm 0.06 \pm 0.02)\%,\\
{\cal{A}}_{CP+}^{DK}& = &( 0.07 \pm 0.18 \pm 0.07)\%,\\
{\cal{A}}_{CP+}^{D\pi}& = &( 0.01 \pm 0.04 \pm 0.01)\%.
\end{eqnarray*}
None of the measured asymmetries significantly deviates from zero, but all results agree with existing measurements within their uncertainties. The main systematic uncertainties are associated to possible differences in the trigger response, to the PID calibration procedure, and to the parameterisation of the background.
With larger data samples, we will be able to extract $\gamma$ and all hadronic unknowns  
by combining ${\cal{A}}_{CP+}^{DK}$ and ${\cal{R}}_{CP+}$ with measurements of additional $\gamma$-sensitive 
observables from other methods.

%%%%%%%%%%%%%%%%%%%%%%%%%%%%%%%%%%
\subsection{\label{ggsz}Towards a GGSZ measurement}

\begin{figure*}[htp!]
\centering
\includegraphics[width=80mm]{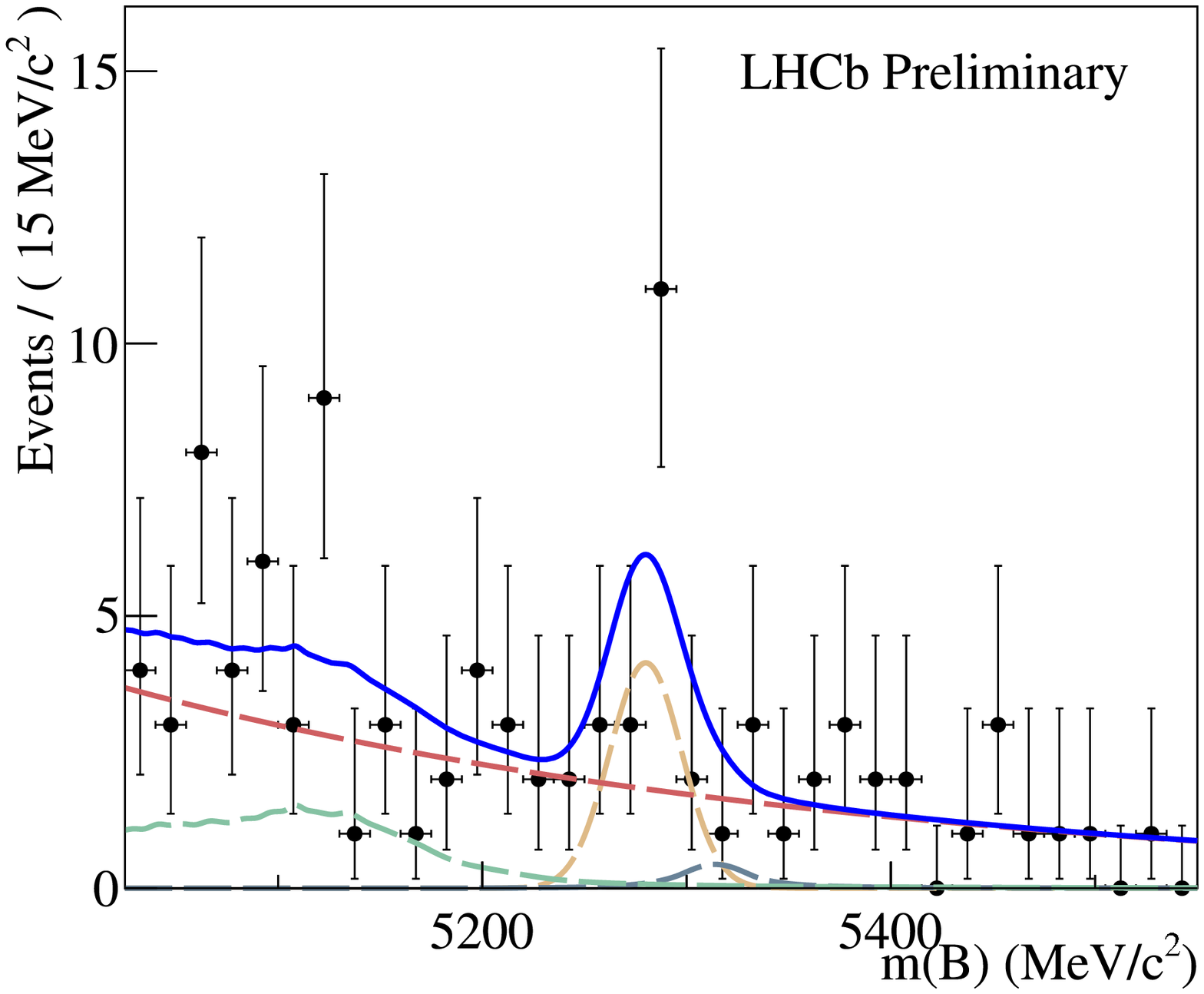}
\includegraphics[width=80mm]{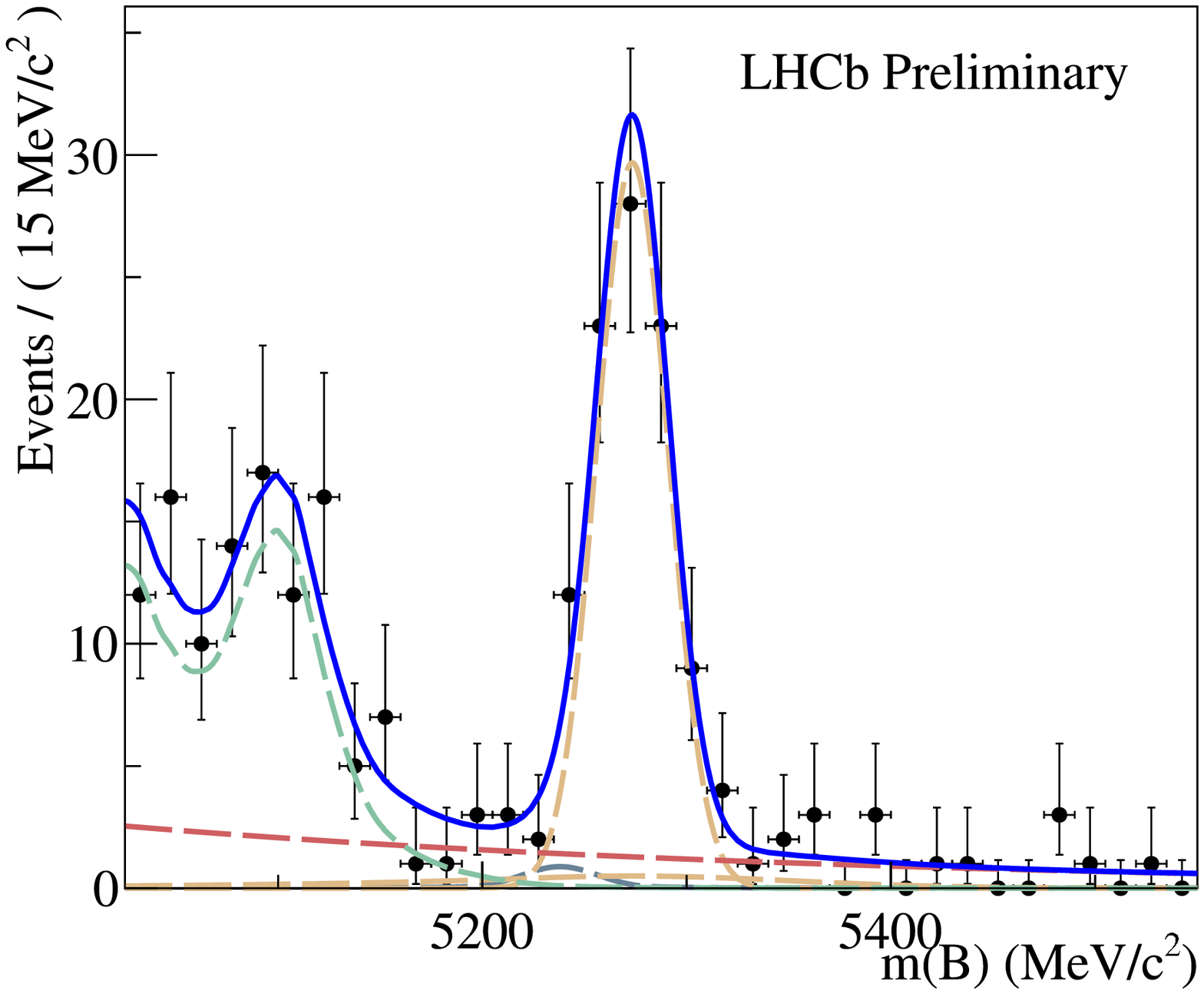}
\caption{The recosntructed $B$ mass distributions for $B^\pm \to (K^0_S\pi\pi)_D K^\pm$ (left) and $B^\pm \to (K^0_S\pi\pi)_D \pi^\pm$ (right). The dashed lines indicate the fitted signal contribution (dark yellow), the background from combinatorial (red) and partially reconstructed decays (light green), and the cross-feed between $B\to DK$ and  $B\to D\pi$ (dark green)\cite{bib:LHCbGLW}.} \label{fig:GGSZ}
\end{figure*}

The ratio of the $B^+\to DK^+$ and  $B^+\to D\pi^+$
branching fractions is also measured in the $D\to K_S^0\pi^+\pi^-$ final state, as it is the first step towards the measurement of $\gamma$ with the GGSZ method~\cite{bib:LHCbGLW}. As for the GLW analysis, the $B^+\to DK^+$ and $B^+\to D\pi^+$ samples are separated by the value of DLL$_{K\pi}$ for the bachelor hadron.
Yields are extracted with a simultaneous fit to the $B$ invariant mass distributions for the two samples. The results are shown in Figure~\ref{fig:GGSZ}.  In 36.5 pb$^{-1}$, the fitted signal yield for $B^+\to D\pi^+$ is 95$^{+14}_{-12}$ events, and the ratio of branching fraction of $B^+\to DK^+$ and $B^+\to D\pi^+$ is
$${\cal{R}}_{K_S^0\pi\pi}^{K/\pi} = (12.0^{+6.0}_{-5.0} \pm 1.0)\%,$$ 
where the largest systematic uncertainty is due to the appropriateness of the fit model (7\%).

%%%%%%%%%%%%%%%%%%%%%%%%%%%%%%%%%%
\subsection{\label{ads}Hunting for the ADS suppressed modes}

The $CP$ asymmetry in the decay rates of the suppressed ADS mode $B^+\to D(K^-\pi^+) K^+$ is expected to be enhanced by the fact that the two interfering amplitudes in these modes have similar size. However, the branching fraction of this mode is small, ${\cal{O}}(10^{-7})$, hence its observation is difficult. The most competitive result to date is from the Belle Collaboration~\cite{bib:BelleADS}, who observed $56.0^{+15.1}_{-14.2}$ events ($4.1~\sigma$ significance) in their full data set of $772\times 10^6$ $B\bar{B}$ pairs collected at the $\Upsilon(4S)$.

\begin{figure}[ht]
\centering
\includegraphics[width=80mm]{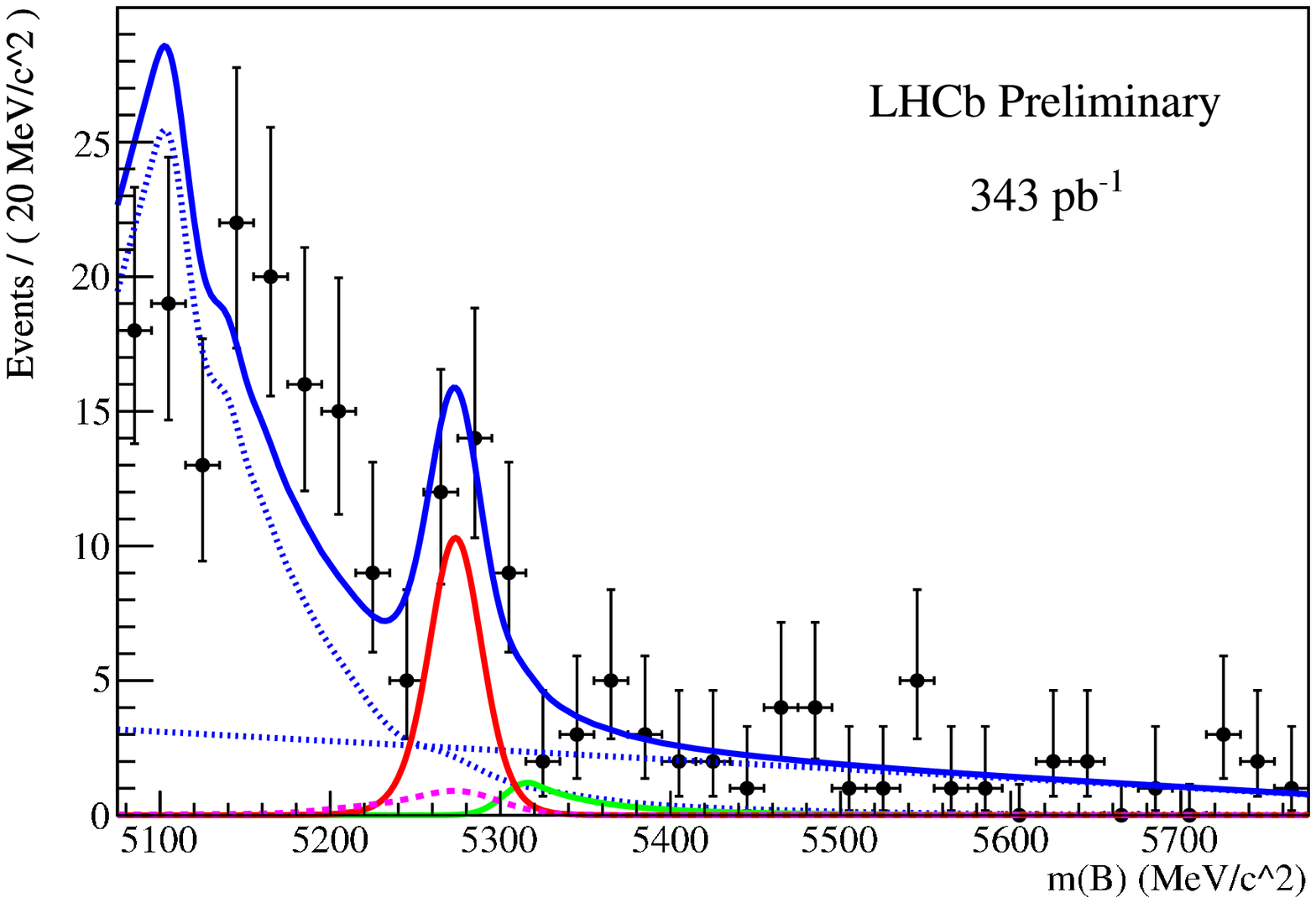}
\caption{The reconstructed $B$ mass distribution for the ADS suppressed candidates, summed over both $B$ charges. The red line indicates the signal component. The background components are from combinatorial and partial reconstruction (dashed blue), charmless sources (magenta), and $B^\pm\to D\pi^\pm$ (green)\cite{bib:LHCbADS}.
} \label{fig:ADS}
\end{figure}

The LHCb collaboration has performed a search for these modes using the early 2011 data-set, corresponding to 343 pb$^{-1}$ of data~\cite{bib:LHCbADS}. The analysis is similar to the GLW analysis just described. An improved event selection, based on a ``Boosted Decision Tree'' algorithm, is used in this case to isolate the $B^+ \to D(K\pi) h^+, h={K,\pi}$ candidates from the background. As for the previously described analyses, the value of the particle identification variable DLL$_{K\pi}$ for the $B$-meson bachelor track is used to effectively separate $B\to D\pi$ from $B\to DK$. 
Eight signal yields are extracted with an unbinned maximum-likelihood fit, corresponding to the two $B$-meson charges, the two product of kaon charges (opposite-sign kaons are suppressed and same-sign kaons are favoured), and the two fail/pass slices according to the PID requirement on the bachelor, DLL$_{K\pi}>$4.  Particular attention has been paid to model the signal and the different background components in the fit. The results of the fit to the suppressed modes, summed over both $B$ charges, are shown in Figure~\ref{fig:ADS}. 
The charge asymmetry between the $B^-$ and $B^+$ suppressed modes is measured to be
$$A_{ADS}^{DK} = -0.39 \pm 0.17 \pm 0.02,$$ 
and the average partial rate $R_{ADS}^{DK}$ of the suppressed over favoured mode
$$R_{ADS}^{DK} = (1.66 \pm 0.39 \pm 0.24) \times 10^{-2},$$
which corresponds to $4.0~\sigma$ significance for the evidence of the suppressed decay.
 The main sources of systematic uncertainties are the PID calibration procedure and the background model.
All these preliminary results are highly competitive with existing measurements and consistent with world averages.

%%%%%%%%%%%%%%%%%%%%%%%%%%%%%%%%%%
\section{\label{strange}Measurements of the {\boldmath $B^0_s$} production fraction} 

Strange $B$ mesons offer a still largely unexplored window on $CP$ violation studies and searches of physics beyond the Standard Model. For example, the determination of the absolute branching fraction for the rare decay $B^0_s\to\mu^+\mu^-$ provides an important constraint to different new physics models. The LHCb measurement of ${\cal{B}}(B^0_s\to\mu^+\mu^-)$~\cite{bib:Olivier} and of other $B^0_s$ decay branching fractions relies on the knowledge of $f_s/f_d$, the ratio of $B^0_s$ production to $B^0$ production.

We have performed two measurements of the ratio $f_s/f_d$ using the relative abundance of $B^0_s\to D^-_s\pi^+$ to $B^0\to D^-K^+$, and to $B^0\to D^-\pi^+$ decays, and a measurement of the ratio $f_s/(f_u + f_d)$, where $f_u$ is the $B^+$ production fraction, using $H_b$ semileptonic decays, identified by the detection of a muon and a charmed hadron. In this section, we report on the hadronic~\cite{bib:hadronic} and semileptonic~\cite{bib:semileptonic} measurements, and on their combination~\cite{bib:average}.

\subsection{\label{hadronic}{\boldmath $f_s/f_d$} from hadronic decays} 

The reconstruction of $B^0_s\to D_s^- \pi^+$ decays is the first step towards the time-dependent analysis of $B^0_s\to D_s^- K^+$, which is sensitive to $\gamma$. In addition, the ratio of its branching fraction to U-spin related $B^0$ decay modes can be used to measure $f_s/f_d$~\cite{bib:FST}.
Here, we report on the latter measurement, which has been performed with a sample of 35 pb$^{-1}$ collected in 2010. Two normalisation modes are used: $B^0\to D^-K^+$ and $B^0 \to D^-\pi^+$. The first is dominated by contributions from colour-allowed tree-diagram amplitudes, and is therefore theoretically well-understood. The second leads to a smaller statistical uncertainty due to its greater yield, but suffers from an additional theoretical uncertainty due to the contribution from a $W$-exchange diagram.

\begin{figure}[ht]
\centering
\includegraphics[width=52mm]{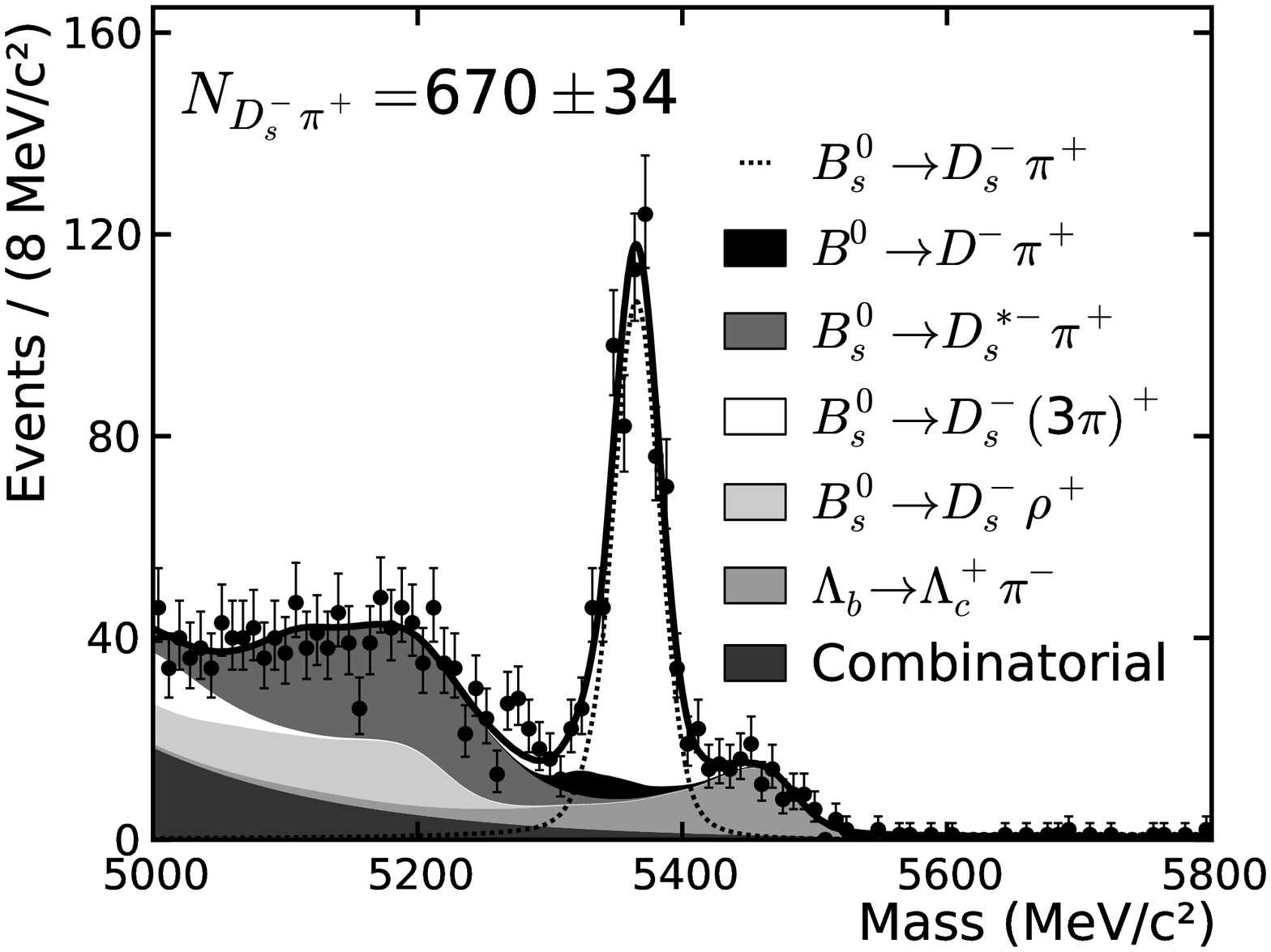}
\includegraphics[width=52mm]{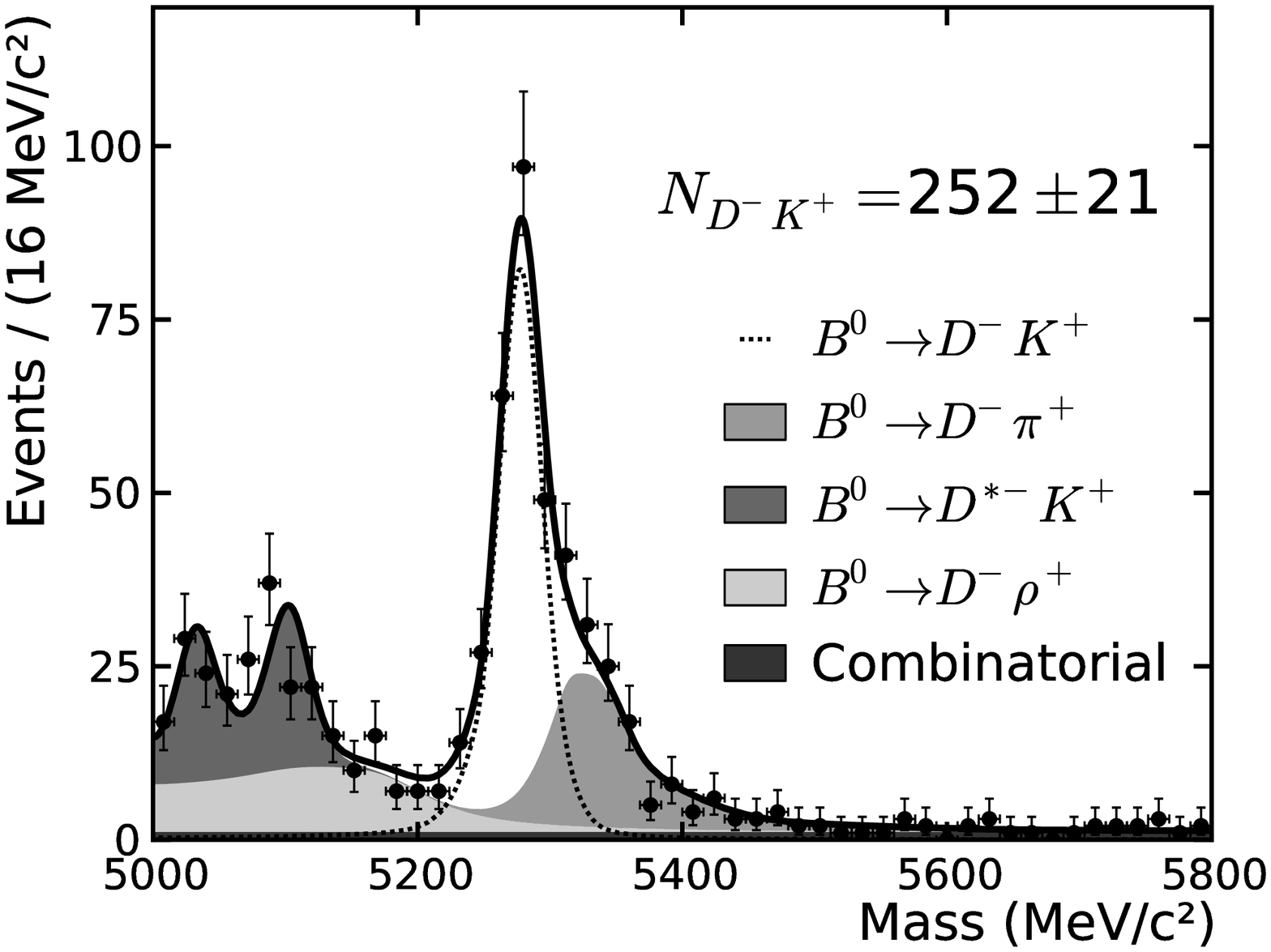}
\includegraphics[width=52mm]{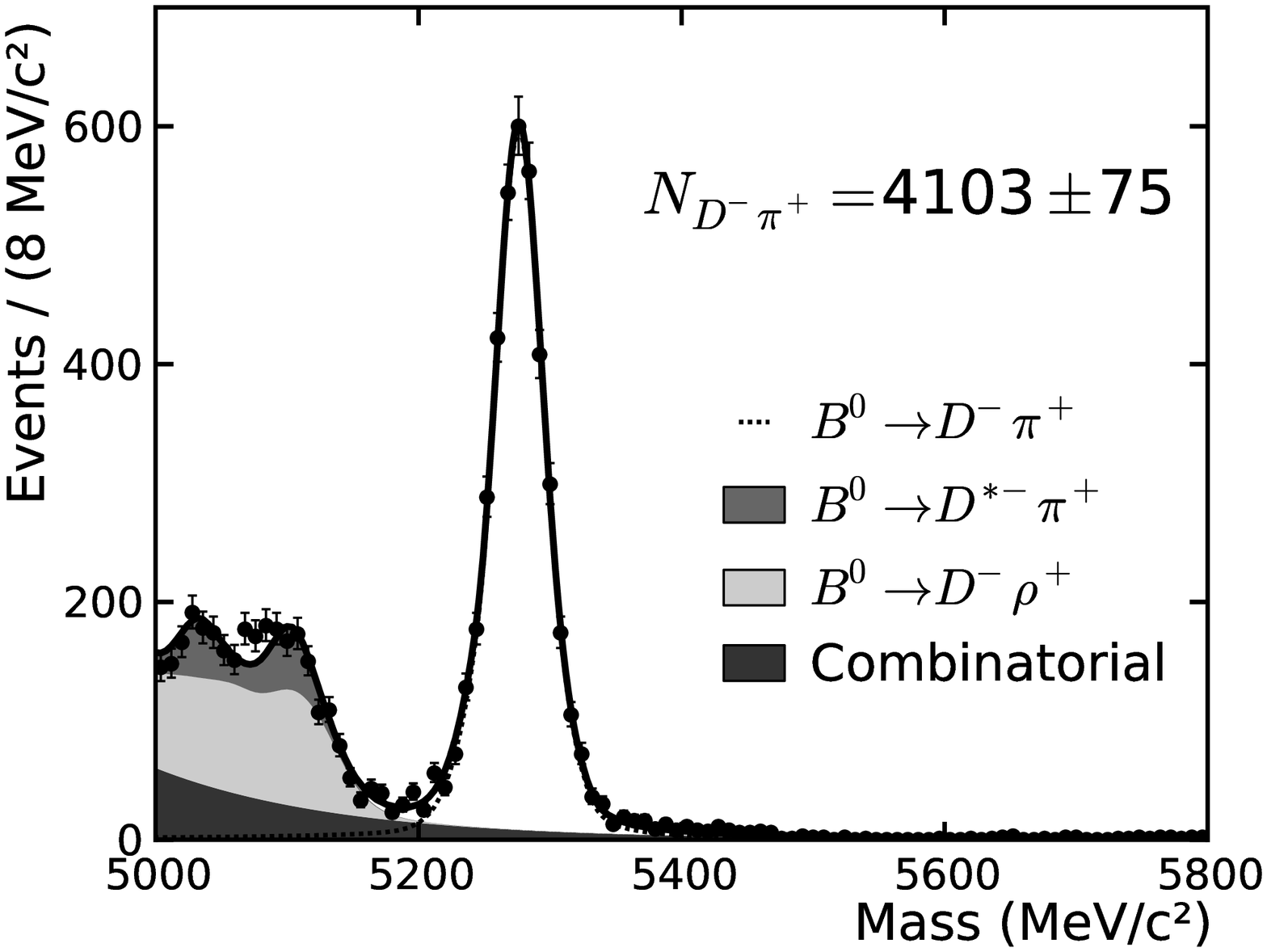}
\caption{Mass distributions of the $B^0_s\to D_s^- \pi^+$, $B^0\to D^-K^+$, and $B^0 \to D^-\pi^+$ candidates (left to right)\cite{bib:hadronic}.} \label{fig:hadronic}
\end{figure}

The relative yields of the three decay modes are extracted from unbinned maximum likelihood fits to the mass distributions, which are shown in Fig.~\ref{fig:hadronic}.

The value of $f_s/f_d$ is found to be 
$$f_s/f_d = 0.250 \pm 0.024 {\mathrm{(stat.)}} \pm 0.017 {\mathrm{(syst.)}} \pm 0.017 {\mathrm{(theor.)}}$$ from the relative yields of $B^0_s\to D_s^- \pi^+$ with respect to $B^0\to D^-K^+$, and
$$f_s/f_d = 0.256 \pm 0.014 {\mathrm{(stat.)}} \pm 0.019 {\mathrm{(syst.)}} \pm 0.026 {\mathrm{(theor.)}}$$ 
from $B^0_s\to D_s^- \pi^+$ with respect to $B^0\to D^-\pi^+$.

\subsection{\label{semileptonic}$b$-hadron production fraction measurements from semileptonic decays} 

The semileptonic measurement of the $b$-hadron production fractions is based on 3 pb$^{-1}$ of LHCb data collected in 2010. We measure two production ratios: that of $\bar{B^0_s}$ and that of $\Lambda_b^0$  relative to the sum of $B^-$ and $\bar{B^0}$. The relative fractions are extracted from the yields in four different final states: $D^0\mu^-\bar\nu X$, $D^+\mu^-\bar\nu X$, $D_s\mu^-\bar\nu X$, and $\Lambda_c\mu^-\bar\nu X$. We do not attempt to separate $f_u$ and $f_d$, but we measure their sum from $D^0$ and $D^+$ channels, taking into account corrections due to cross-feed from $\bar{B^0_s}$ and $\Lambda_b^0$ decays.

\begin{figure}[ht]
\centering
\includegraphics[width=135mm]{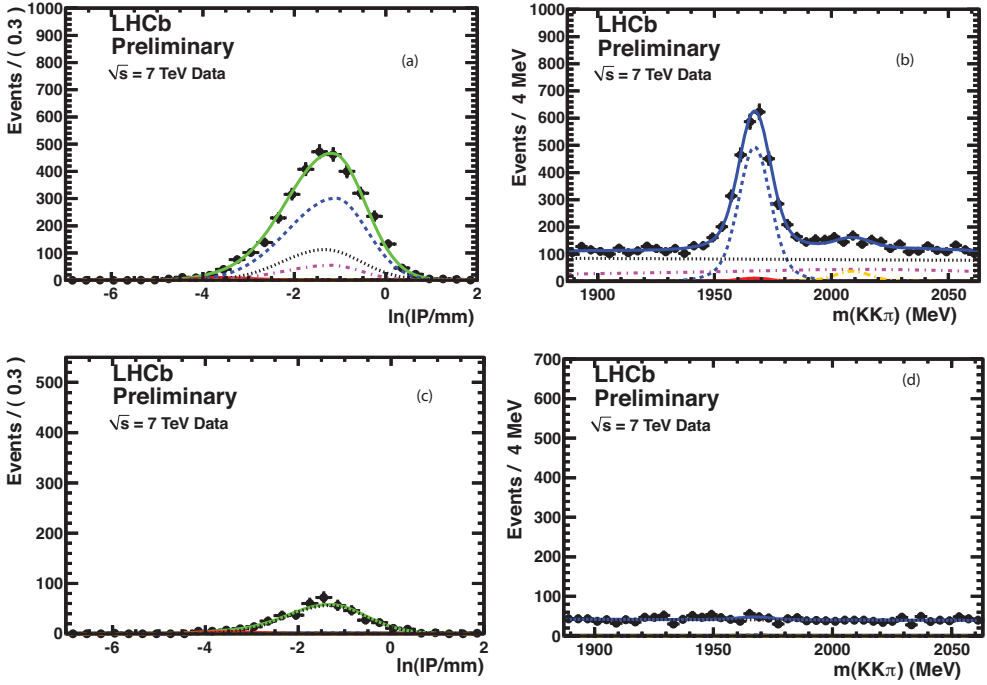}
\caption{The logarithm of the IP distributions for (a) right sign and (c) wrong sign $D^0$ candidate combinations with a muon. The dotted curves show the combinatorial backgrounds, the small red-solid curve the prompt-charm contributions, the dashed curves the signal, the purple-dashed curves represent a background originating from $\Lambda_c$ reflection, and the green-solid curves the total. The invariant $K^-K^+\pi^+$ mass spectra are also shown for right sign (b) and wrong-sign (d) combinations\cite{bib:semileptonic}.} \label{fig:semileptonic}
\end{figure}

The $H_b$ signals are separated from various sources of background yields by studying the two-dimensional distributions of the charm candidate invariant mass and impact parameter (IP) with regard to the primary $pp$ collision vertex. 
This approach allows us to determine the background coming from false combinations and from prompt charm production.
As an example, the results of the fit for the $D^+_s\mu^-\bar\nu X$ candidates are shown in Fig.~\ref{fig:semileptonic}.
The fractions $f_s/(f_u + f_d)$ and $f_{\Lambda_b^0}/(f_u + f_d)$ are determined as function of the pseudo-rapidity $\eta$, and the charmed hadron-muon pair transverse momentum, $p_t$. 
\begin{figure}[ht]
\centering
\includegraphics[width=80mm]{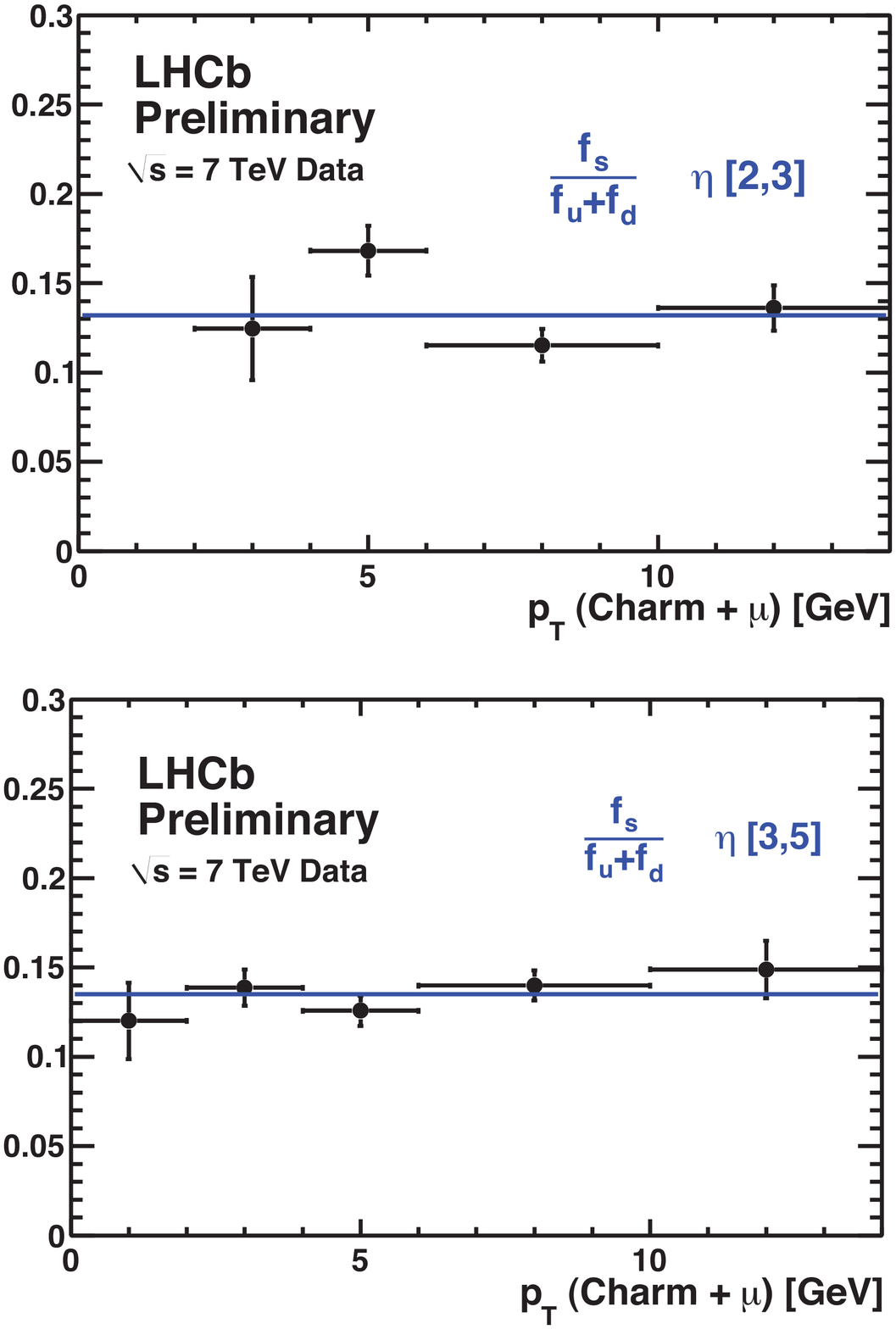}
\caption{Ratio between $B^0_s$ and light $B$ meson production fractions as a function of the transverse momentum of the $D_s\mu$ pair in two bins of $\eta$. The errors shown are statistical only
\cite{bib:semileptonic}.} \label{fig:fragmentation}
\end{figure}
We find
$$ f_s/(f_u + f_d) = 0.134 \pm 0.004 ^{+0.011}_{-0.010}, $$
with no significant dependence on $\eta$, nor $p_t$, as shown in Fig.~\ref{fig:fragmentation}. The main systematic uncertainty is from limited knowledge of the charm-hadron branching fractions.
A dependence on $p_t$ is instead found for the $\Lambda_b^0$ fragmentation function with regard to the sum of $B^-$ and $\bar{B^0}$. Assuming a linear dependence, we get $f_{\Lambda_b}/(f_u + f_d) = (0.404 \pm 0.017 \pm 0.027 \pm 0.105) \times [ 1-(0.031 \pm 0.004 \pm 0.003) \times p_t ({\mathrm{GeV}})]$, where the errors are statistical, systematic and (for the constant term) an absolute scale uncertainty due to the error in ${\cal{B}}(\Lambda_c\to pK\pi)$, respectively. No $\eta$ dependence is found. More details on this measurement can be found in Ref.~\cite{bib:semileptonic}.

\subsection{Average}

If we use isospin symmetry to set $f_u = f_d$, the LHCb measurements of the ratio of strange $B$ meson to light neutral $B$ meson, obtained using $H_b$ semileptonic decays, is in good agreement with the two measurements obtained with hadronic decays. Therefore, we combine them to derive 
$$ f_s/f_d = 0.267^{+0.021}_{-0.020}. $$
Since we do not observe a dependence upon properties such as the transverse momentum or rapidity of the $B$ meson, although our measurement is obtained from data within the LHCb acceptance, it is reasonable to assume that it is valid in other phase-space regions.
We also note that, despite the fact that this ratio is not a-priori universal, our result is in remarkable agreement with the average between results from LEP and Tevatron experiments ($f_s/f_d = 0.271\pm 0.027$~\cite{bib:HFAG}).

%%%%%%%%%%%%%%%%%%%%%%%%%%%%%%%%%%
\section{\label{baryons}Studies of beauty baryon decays to charm}

The study of $b$-baryons is a largely unexplored area where LHCb has great potential for measurements of spectroscopy and $CP$ violation. In particular, we look for $\Lambda_b^0 \to D p K$, which is an unobserved $\Lambda_b^0$ decay mode. This channel is sensitive to the angle $\gamma$~\cite{bib:DpK}, similarly to $\Lambda_b^0 \to D \Lambda$, as originally proposed in Ref.~\cite{bib:DLambda}, but it presents some advantages compared to $D\Lambda$, because the $pK$ pair originates from the $\Lambda_b^0$ decay vertex, rather than from a long-lived intermediate particle, therefore a larger reconstruction efficiency is expected at LHCb. In addition, the use of the full phase-space of the three-body decay may enhance the sensitivity compared to the two-body process.

\begin{figure}[htp!]
\centering
\includegraphics[width=80mm]{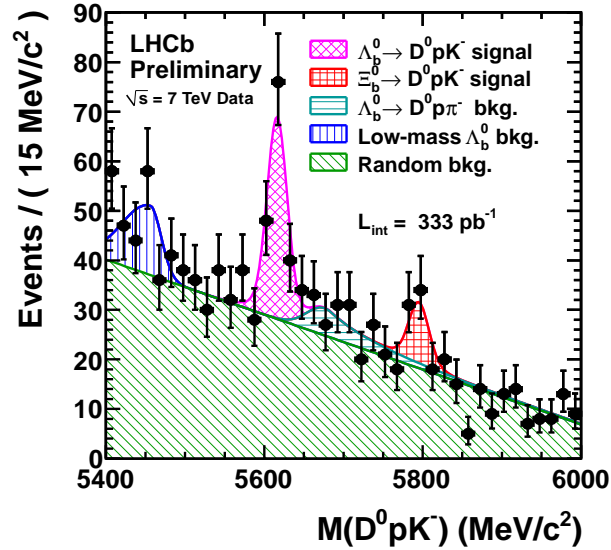}
\caption{Invariant mass spectrum of $DpK^-$. Results of the fit are overlaid to the data points\cite{bib:DpK}.} \label{fig:DpK}
\end{figure}

The $D^0pK^-$ final state is studied together with the $\Lambda_b^0\to D^0p\pi^-$, and $\Lambda_b^0\to \Lambda_c^+\pi^-$ decays, which have similar kinematics and can be used as normalisation channels. With 333 pb$^{-1}$ taken by LHCb in early 2011, we measure the ratio of branching fractions
$$ \frac{{\cal{B}}(\Lambda_b^0\to D^0p\pi^-)\times{\cal{B}}(D^0\to K^- \pi^+)} {{\cal{B}}(\Lambda_b^0\to \Lambda_c^+ \pi^-)\times{\cal{B}}(\Lambda_c^+\to K^- p\pi^+)} = 0.119 \pm 0.006 \pm 0.013.$$
We also present the first observation of the $\Lambda_b^0 \to D p K$ decay and measure the ratio of branching fractions
$$ \frac{{\cal{B}}(\Lambda_b^0\to D^0pK^-)}{{\cal{B}}(\Lambda_b^0\to D^0 p\pi^-)} = 0.112 \pm 0.019 ^{+0.011}_{-0.014}.$$
The significance of the $\Lambda_b^0 \to D p K$ signal is 6.3~$\sigma$. As seen in Fig.~\ref{fig:DpK}, in the $DpK^-$ final state we find a hint of production of the neutral beauty-strange baryon $\Xi^0_b$ with significance of 2.6~$\sigma$, and we measure the ratio of branching fraction times production ratio with respect to those for the $\Lambda_b^0$ 
$$ \frac{f_{b\to\Xi^0_b}\times{\cal{B}}(\Xi_b^0\to D^0pK^-)}{f_{b\to\Lambda^0_b}\times{\cal{B}}(\Lambda_b^0\to D^0pK^-)}= 0.29 \pm 0.12 \pm 0.18.$$
We measure the difference of the $\Xi^0_b$ and $\Lambda_b^0$ masses to be equal to (181.8 $\pm$ 5.5 $\pm$ 0.5) MeV/c$^2$, which is in good agreement with the recent measurement of the CDF Collaboration~\cite{bib:XiCDF}.%, who established the esistence of this state in a different decay mode ($\Xi_b^0\to\Xi_c^+\pi^-$).
The main sources of systematic uncertainty in these results are the description of the signal and background lineshapes, and, for the branching fraction measurements, the determination of reconstruction and PID efficiency ratios.

\section{\label{conclusions}Conclusions}

LHCb is on track for a precise measurement of the CKM angle $\gamma$ using $b$-hadron decays to open charm, with both well-established modes (GLW, ADS, GGSZ) and unique ways (e.g., using $B^0_s\to D^\pm_s K^\mp$, and $\Lambda_b^0\to DpK$ decays). In particular,
the 4.0~$\sigma$ evidence of the ADS suppressed mode is highly competitive with the previous measurements.

Another important by-product of the study of $b$-hadron decays to open charm is the most precise measurement of $f_s/f_d$, the 
$B^0_s$ production fraction with respect to that of $B^0_d$, from the combination of LHCb results obtained with hadronic and semileptonic decays.

All these results have been obtained  with data samples from 2010 or early 2011, which are just a fraction of the total expected by the end of this year (1 fb$^{-1}$). Hence, an improved precision in the measurement of $\gamma$ and many other measurements with {\it{charming}} tree-level final states can be expected very soon from LHCb.

\begin{acknowledgments}
I am grateful to Steven Blusk, Tim Gershon, Vava Gligorov, and Olaf Steinkamp for the help in the preparation of this contribution. I would like also to thank the organisers of DPF2011 for inviting us to such an exquisite conference.
\end{acknowledgments}

\bigskip % extra skip inserted

\end{document}